\documentclass[]{USG}
\usepackage{anyfontsize} %
\usepackage{colortbl}   
\usepackage{xcolor}     
\usepackage{steinmetz}
\graphicspath{{./images/}}
\articletype{ORIGINAL ARTICLE}%
\journal{}
\volume{0}
\copyyear{2026}
\startpage{1}
\articledoi{10.1002/0000}

\usepackage{graphicx}%
\usepackage{multirow}%
\usepackage{amsmath,amssymb,amsfonts}%
\usepackage{amsthm}%
\usepackage{mathrsfs}%
\usepackage{textcomp}%
\usepackage{manyfoot}%
\usepackage{booktabs}%
\usepackage{algorithm}%
\usepackage{algorithmicx}%
\usepackage{algpseudocode}%
\usepackage{listings}%
\usepackage{geometry}
\usepackage{enumitem}
\usepackage{placeins}
\usepackage{multirow}
\usepackage{bm}
\usepackage{amsmath}
\usepackage{caption}
\usepackage{float}
\usepackage{tikz,pgfplots}
\pgfplotsset{compat=1.17}
\usetikzlibrary{positioning}
\usetikzlibrary{arrows}
\usepackage{NJDnatbib}

\begin{document}


\title{Log-F-penalized Conditional Logistic Regression for Sparse Data}
\transtitle{Log-F-penalized Conditional Logistic Regression for Sparse Data}
\subtranstitle{trans-subtitle}
\author[1]{Ying Yu}
\author[2]{Jiying Wen}
\author[3]{Jinko Graham}
\author[4]{Brad McNeney}
\authormark{YU \textsc{et al.}}
\titlemark{Log-F-penalized Conditional Logistic Regression for Sparse Data}

\address[1]{\orgdiv{Department of Statistics and Actuarial Science, Simon Fraser University, Burnaby, BC, Canada} }\orgname{Simon Fraser University, }%
\orgaddress{\state{British Columbia, }\country{Canada}}

\corres{Brad McNeney  (\email{mcneney@sfu.ca}) ~|~ Ying Yu (\email{daisy0130@gmail.com})}

\fundingInfo{This work was supported by Discovery Grant RGPIN/05595-2019 to BM and RGPIN/04296-2018 to JG from the Natural Sciences and Engineering Research Council of Canada (NSERC), and used the computational resources provided by the Digital Research Alliance of Canada \url{https://www.alliancecan.ca/en}}


\abstract{We investigate penalized likelihood methods for estimation and inference in conditional logistic regression. The standard conditional maximum likelihood estimator is known to be biased away from zero in small or sparse matched case-control studies. A widely used remedy is Firth's penalized likelihood approach, which has good frequentist operating characteristics but provides limited control over the degree of shrinkage applied to individual regression coefficients.
We develop point and interval estimators by penalizing the conditional likelihood with independent log-$F$ distributions.
The log-\(F\)-penalized approach allows analysts to calibrate  shrinkage using interpretable prior assumptions about plausible effect sizes. We also provide practical guidance for calibrating the amount of shrinkage and show that the method can be implemented through data augmentation using standard conditional logistic regression software.
We illustrate the methods using data from (i) a study of maternal exposure to diethylstilbestrol and the risk of vaginal cancer in daughters, and (ii) a genetic association study of type 2 diabetes. We then compare the log-$F$-penalized approach with Firth's penalized likelihood method in a simulation study.
In simulations, the log-$F$-penalized estimators had
confidence-interval coverage comparable to that of Firth's method and
lower mean squared error, with similar type~1 error rates and power.
These results support the use of log-$F$-penalized conditional logistic regression for inference in sparse matched and stratified studies.}




\maketitle

\section{Introduction} \label{Intro}

Conditional logistic regression is widely used to analyze binary outcomes from subjects sampled in strata or matched sets \citep{HeinzePuhr10}. In this setting, the regression parameters represent log-odds ratios. Conditioning on sufficient statistics for stratum- or matched-set-specific nuisance parameters eliminates them from the likelihood, enabling consistent inference for the regression parameters of interest \citep{BreslowDay80}. Matching is particularly valuable when a measured variable, such as school or family, serves as a surrogate for unmeasured confounders like environmental exposures, socioeconomic status, or genetic ancestry \citep{Mansournia18}. In such cases, directly adjusting for all confounders is not feasible, and conditional logistic regression becomes the only practical option \citep{Brumback2012}.

Conditional logistic regression also arises in contexts beyond stratified or matched case-control studies, such as genetic analyses of affected children and their parents. Case-parent trio studies, for example, condition on parental genotypes \citep{Schaid93}. For a given genetic marker, conditioning creates a matched set consisting of the alleles transmitted to the affected child (the ``case'') and the possible alleles that could have been transmitted (the ``controls''). Here, the regression parameters correspond to log-genotype relative risks, comparing disease risk between two genotypes at a marker while holding covariates fixed.

Sparse data provide limited support for estimating model parameters. With a categorical exposure, sparseness arises when few cases or controls exist at some exposure–outcome combinations \citep{greenland2016sparse}. Sparseness is therefore more likely for rare exposures than for common exposures. With a continuous exposure, matched sets are less informative when cases and controls have similar exposure values. Under sparse data, the conditional maximum likelihood estimator (CMLE) of regression coefficients is biased away from zero \citep{Greenland00a,Greenland00b,HeinzePuhr10}. The extent of this bias depends on features such as the number of matched sets, the amount of exposure variation between cases and controls within matched sets, and the number of covariates included in the model.

\begin{figure}
\centering
\includegraphics[width=0.80\linewidth]{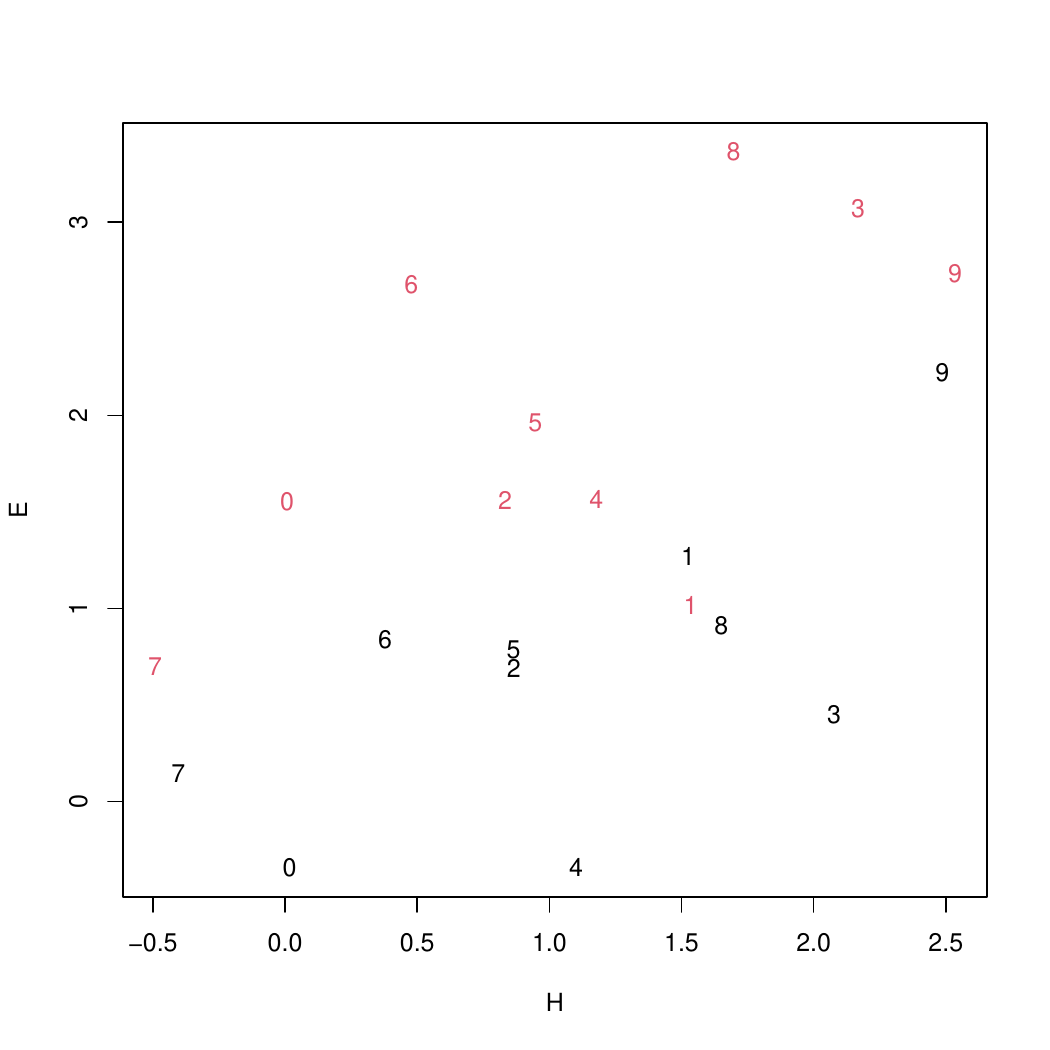}
\caption{Simulated data on a continuous exposure $E$ versus a confounding variable $H$ for 10 matched sets under 1:1 matching. Points are labelled by matched-set number, with cases in red and controls in black font. The confounding variable is included in the display only to offset the matched sets horizontally. The exposure value of about 1.5 nearly separates cases from controls.}
\label{fig:exdatplot}
\end{figure}

Bias becomes infinite under \emph{separation}, when a linear combination of covariates perfectly distinguishes cases from controls \citep{AlbertAnderson84}. This phenomenon occurs in the study of \cite{herbst1971adenocarcinoma}, which first identified maternal treatment with diethylstilbestrol (DES) as a risk factor for vaginal cancer in daughters. The matched case-control design paired patients with four hospital-based controls on birth date and room type. All cases were exposed to DES and all controls were unexposed (Table~\ref{DESdat}), resulting in complete separation and an infinite CMLE. Standard conditional logistic regression software fails to converge in such situations. Near separation, where a linear combination almost but not perfectly separates cases from controls, also induces marked bias. Figure~\ref{fig:exdatplot} illustrates near separation in simulated matched sets and a continuous exposure.

\begin{table}
\normalsize
  \centering
  \caption{DES exposure in cases and controls from Herbst {\it et al.}}
  \begin{tabular}{ccc}
  \textbf{Exposure} & \textbf{Cases} & \textbf{Controls} \\ \hline
   yes & 7 & 0  \\
   no  & 1 & 32
  \end{tabular}
  \label{DESdat}
\end{table}

For matched pairs, simple data-augmentation methods can reduce sparse-data bias and avoid infinite estimates. Table~\ref{tab:DES} shows DES exposure among matched pairs derived from the DES data of Herbst \emph{et al.} by selecting one of the four matched controls at random. In the Table, each cell is augmented by a fixed constant $k$ \citep{Bishop75}. When $k=0$ (no augmentation) the CMLE of the odds ratio is infinite. With $k=1/2$ (Haldane’s method) the estimated odds ratio is $(7.5/0.5)=15$; with $k=1$ (Laplace’s method) the estimate is $8/1=8$ \citep{Greenland00b}.

\begin{table}[h]
\centering
\caption{Augmented DES exposure status for eight matched pairs}
\normalsize
\begin{tabular}{cccc}
        & & \multicolumn{2}{c}{Control}  \\
        & & exposed & unexposed \\
        Case  & exposed & $0+k$ & $7+k$ \\
              & unexposed & $0+k$ & $1+k$
\end{tabular}
\label{tab:DES}
\end{table}

\begin{figure}
\centering
\includegraphics[width=0.8\linewidth]{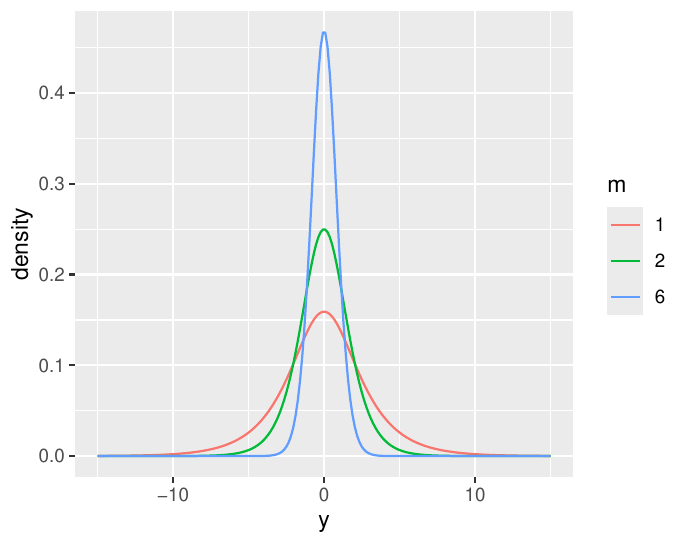}
\caption{Comparison of log-$F(m,m)$ densities for values $m=1$, 2 and 6. The log-$F(m,m)$ distribution is symmetric and centered at zero, with a variance that decreases with increasing $m$.}
\label{fig:log-F densityplot}
\end{figure}

Penalized-likelihood methods maximize the product of the likelihood and a penalty term derived from a prior distribution, to shrink estimators toward the prior mean \citep{greenland2015penalization}.  The CMLE is biased away from zero, and so a zero-centred prior is a sensible strategy. We show in Appendix~\ref{HLlogF} that both Haldane’s and Laplace’s corrections arise from a penalized conditional likelihood with log-$F$ priors. 
A log-$F(m,m)$ random variable is the logarithm of an $F$ distribution with $m$ numerator and denominator degrees of freedom \citep{johnson_kotz_balakrishnan_1994}. \cite{greenland2015penalization} proposed log-$F$-penalized likelihood for unconditional logistic regression, implemented via data augmentation with $m$ pseudo-observations per coefficient. The choice of $m$ can reflect prior beliefs about plausible effect sizes. Figure~\ref{fig:log-F densityplot} illustrates how log-$F$ densities become more concentrated about zero as $m$ increases. Larger values of $m$ induce stronger shrinkage to zero. 

The penalized-likelihood approach of \cite{Firth93} is based on Jeffreys' prior \citep{Jeffreys46}, which defines the penalty through the information matrix.
Firth showed that this penalty removes the first-order asymptotic bias of maximum-likelihood estimators and provides stable estimation in sparse-data settings where conventional maximum-likelihood estimation may be biased or fail because of separation. For logistic regression, the induced Jeffreys prior is a proper, symmetric distribution centered at zero. \cite{HeinzePuhr10} applied Firth's approach to \emph{conditional} logistic regression and demonstrated improved bias and confidence-interval coverage relative to unpenalized conditional logistic regression, exact conditional logistic regression, and penalized unconditional methods. However, because the Jeffreys penalty is determined by the observed information matrix, the amount of shrinkage is implicit and data-dependent rather than controlled by a user-specified parameter with a direct interpretation.

Despite important progress on bias reduction for logistic models, a gap remains 
for matched and stratified study designs. Existing remedies are either limited to studies of a single binary exposure and matched pairs (Haldane and Laplace corrections), or rely on the data-dependent Jeffreys penalty \citep{HeinzePuhr10}. 
Neither approach allows analysts to specify the degree of shrinkage through an interpretable parameter that reflects plausible effect sizes.

In this paper, we extend log-$F$-penalized likelihood methods to conditional logistic regression for 1{:}M matched designs. 
We show that the log-$F(m,m)$ penalty can be implemented by adding a small number of artificial exposure-discordant matched sets, allowing standard conditional logistic regression software to be used without specialized programming. We also provide practical guidance for calibrating the penalty parameter $m$ from scientifically meaningful covariate contrasts and plausible prior odds-ratio ranges. For binary exposures, we follow the recommendations of \cite{greenland2015penalization}, using $m=1$ for an uninformative prior and $m=2$ for a weakly informative prior. For continuous exposures, following \cite{Gelman2008}, we calibrate the prior to a two-standard-deviation change in the exposure variable and discuss how this calibration informs the choice of $m$.

We illustrate log-$F$-penalized conditional logistic regression using data from a matched case-control study and a case-parent trio study. 
We also compare log-$F$-penalized conditional logistic regression with unpenalized and Firth-penalized conditional logistic regression in a simulation study. Exact and approximate exact conditional logistic regression are not considered because previous investigations indicate they have increased bias and lower power than Firth-penalized conditional logistic regression \citep{HeinzePuhr10,GrahamMcNeneyPlatt}.

\section{Methods} \label{Method}

We consider $1{:}M$ matched case–control (or stratified) designs with one case per stratum and possibly varying numbers of controls. We model these data using conditional logistic regression and introduce log-$F(m,m)$ priors on the regression coefficients to obtain penalized estimators. These estimators can be computed with conditional logistic software by augmenting the data with artificial matched sets.

\subsection{Conditional Logistic Regression} \label{CLR}

Suppose that the population has $I$ population strata indexed by $i\in(1,...,I)$ and that, from each stratum, one case and $M_i$ controls are sampled. Let $j\in(0,...,M_i)$ index subjects within each sample stratum, with the case having index $j=0$ and the controls having indices $j=1\ldots,M_i$. Let $\bm{X}_j^i=(X_{j1}^i,\ldots,X_{jK}^i)$ denote the random covariate vector for the $j$th individual in the $i$th stratum, and $\bm{x}_j^i=(x_{j1}^i,\ldots,x_{jK}^i)$ denote the vector of observed values. The regression coefficients are denoted $\bm{\beta}=(\beta_1,...,\beta_K)^T$.
From equation (7.2) of \cite{BreslowDay80}, the conditional likelihood and its log are respectively:
\begin{align}
   L(\bm{\beta})&=    \prod_{i=1}^{I}\frac{\exp(\bm{x}_0^{iT}\bm{\beta})}{\sum_{j=0}^{M_i}\exp(\bm{x}_j^{iT}\bm\beta)} \quad \mbox{   and   }
   \label{cond_lkhd}
\end{align}
\begin{align}
   l(\bm{\beta})&=\sum_{i=1}^{I}\left[\bm{x}_0^{iT}\bm{\beta}-\log\left(\sum_{j=0}^{M_i}\exp(\bm{x}_j^{iT}\bm\beta)\right)\right].
   \label{cond_log_lkhd}
\end{align}
The conditional maximum-likelihood estimator (CMLE) is the argument that maximizes these equations. 
For future reference, we note that when all subjects in a matched set $i$ have the same value $x_k$ of the $k$th covariate (i.e. $x_{jk}^{i}=x_k$ for $j=0\ldots M_i$) the corresponding regression coefficient $\beta_k$ cancels out of the likelihood contribution for the matched set. Therefore, the matched set contributes no information to estimation of $\beta_k$.

\subsection{log-$F$ Penalization} \label{LogF}

The regression coefficients $\beta_k$ ($k=1,\ldots,K$) are assigned independent
log-$F(m_k,m_k)$ priors \citep{johnson_kotz_balakrishnan_1994} with density
\[
f(\beta_k\mid m_k)\ \propto\ \frac{\exp\!\left(\tfrac{m_k}{2}\beta_k\right)}{\bigl(1+\exp(\beta_k)\bigr)^{m_k}}, \qquad k=1,\ldots,K.
\]
Setting $m_k=0$ yields a flat prior and thus no penalization. These priors are symmetric about zero and become more concentrated as $m_k$ increases; consequently, larger $m_k$ induces greater shrinkage of the estimator for $\beta_k$ toward zero.
The resulting penalized likelihood is:
\begin{align}
    L^*(\bm{\beta})
    = L(\bm{\beta})\times\prod_{k=1}^{K}f(\beta_k|m_k)
    = \prod_{i=1}^{I}\frac{\exp(\bm{x}_0^{iT}\bm{\beta})}{\sum_{j=0}^{M_i}\exp(\bm{x}_j^{iT}\bm\beta)}\times\prod_{k=1}^{K}\frac{\exp(\frac{m_k}{2}\beta_k)}{(1+\exp(\beta_k))^{m_k}}. \
\end{align}
Taking logs, we obtain the penalized log-likelihood:
\begin{align}
    l^*(\bm{\beta})
    &= \sum_{i=1}^{I}\left[\bm{x}_0^{iT}\bm{\beta}-\log\left(\sum_{j=0}^{M_i}\exp(\bm{x}_j^{iT}\bm\beta)\right)\right]+\sum_{k=1}^{K}\frac{m_k}{2}\left[\beta_k-2\log(1+\exp(\beta_k))\right] .
    \label{eqn:penloglik}
\end{align}
To obtain the log-$F$-penalized estimator, $\hat{\bm{\beta}}$, we solve the $K$ modified score equations:
\begin{align}
    \frac{\partial l^*(\bm{\beta})}{\partial\beta_k}
    = \sum_{i=1}^{I}\left[x_{0k}^{i}-\frac{\sum_{j=0}^{M_i}x_{jk}^{i}\exp(\bm{x}_j^{iT}\bm{\beta})}{\sum_{j=0}^{M_i}\exp(\bm{x}_j^{iT}\bm{\beta})}\right]+\frac{m_k}{2}\left( 1 - 2\frac{\exp(\beta_k)}{1+\exp(\beta_k)}\right) = 0
\end{align}
for $k=1,...,K$.

We adopt the inferential framework of \cite{HeinzePuhr10} throughout.
Standard errors for $\hat{\bm{\beta}}$ can be obtained from the inverse of the modified observed Fisher information
$$
 [I^*(\hat{\beta})]^{-1}=\left[ \left.-\frac{\partial^2 l^*(\bm{\beta})}{\partial\bm{\beta} \partial\bm{\beta}^T} \right|_{\hat{\bm{\beta}}} \right]^{-1}.
$$
Specifically, the standard error for $\hat{\beta}_{k}$ is the square-root of the $k$th diagonal element of $[ I^*(\hat{\beta}) ]^{-1}$.
A likelihood-ratio statistic is based on the profile penalized conditional likelihood 
$$l_p^*(\beta_k) \equiv l^*(\beta_k,\hat{\bm{\beta}}_{-k}(\beta_k)),$$ where $\bm{\beta}_{-k}$ is the vector of all regression coefficients except $\beta_k$, and $\hat{\bm{\beta}}_{-k}(\beta_k)$ is the value of $\bm{\beta}_{-k}$ that maximizes $l^*(\beta_k,\bm{\beta}_{-k})$ holding $\beta_k$ fixed.
The test statistic for testing $H_0:\beta_k=0$ \emph{versus}
$H_1:\beta_k\not= 0$ is
$$\Lambda^*_k = 2[l_p^*(\hat{\beta}_k)-l_p^*(0)],$$
where $\hat{\beta}_k$ is the maximizer of the profile penalized conditional likelihood. Under the null hypothesis,  $\Lambda^*_k$ has an asymptotic chi-square distribution with one degree of freedom. Level-$\alpha$ confidence intervals are based on inverting the test; i.e.,
solving $$2[l_p^*(\hat{\beta}_k)-l_p^*(\beta_k)]=\chi^2_{1,1-\alpha}$$ for $\beta_k$,
where $\chi^2_{1,1-\alpha}$ is the $1-\alpha$ quantile of the chi-squared distribution with one degree of freedom. These are referred to as PPCL or profile, penalized, conditional likelihood intervals.

\subsection{Implementation}\label{Data_aug}

The prior distribution of $\beta_k$ contributes the following to the penalized log-likelihood in equation (\ref{eqn:penloglik}):
\begin{equation}
\frac{m_k}{2}\bigl(\beta_k - 2\log\{1+\exp(\beta_k)\}\bigr).
\label{eqn:penaltycontribk}
\end{equation}
This term can be interpreted as the weighted log-likelihood contribution of two \emph{artificial} matched pairs that are discordant only on the $k$th covariate:

\bigskip

\noindent \textbf{Data Augmentation for $\beta_k$} (Artificial matched pairs) \\ \vspace*{-.2in}
\begin{itemize}
\item \textbf{Pair A$_k$:} the case has $x_k=1$ and all other covariates $0$, while the control has all covariates $0$. From equations~(\ref{cond_lkhd})–(\ref{cond_log_lkhd}), the likelihood and log-likelihood contributions are $\exp(\beta_k)/\{\exp(\beta_k)+1\}$ and $\beta_k-\log\{1+\exp(\beta_k)\}$, respectively.
\item \textbf{Pair B$_k$:} the case has all covariates $0$, while the control has $x_k=1$ and all other covariates $0$. The corresponding contributions are $1/\{1+\exp(\beta_k)\}$ and $-\log\{1+\exp(\beta_k)\}$.
\end{itemize}

\bigskip

\noindent Adding the two log-contributions and multiplying by the weight $m_k/2$ involving the degrees of freedom for the $k$th regression coefficient yields the expression in equation~(\ref{eqn:penaltycontribk}). Summing over $k=1,\ldots,K$ produces the penalty term in equation~(\ref{eqn:penloglik}). Thus, the log-$F(m,m)$ penalized estimator of $\beta_k$ is the CMLE for the \emph{weighted, augmented} dataset that includes these pseudo–matched pairs. Assuming the software allows matched-set specific weights, implementing the penalty for coefficient $\beta_k$ involves adding the artificial matched pairs $\bm{A}_k$ and $\bm{B}_k$ for each $k$ and weighting them by $m_k/2$. If the software does not allow such weights and $m_k$ is even, one alternative is to replicate each matched pair $\bm{A}_k$ and $\bm{B}_k$ in the data $m_k/2$ times. 

In the special case of matched pairs with a single binary exposure, $x_1$, this weighted augmentation reduces to adding $m_1/2$ to each discordant cell of the $2\times 2$ table of case \emph{versus} control exposure status (Appendix~\ref{HLlogF}). In this setting the choices $m_1=1$ and $m_1=2$ reproduce Haldane’s and Laplace’s corrections, respectively \citep{Greenland00b}.

\subsection{Choice of $m$}
\label{sec:choosem}

Our discussion of the choice of $m$ follows \cite{greenland2015penalization}.
As shown in Figure~\ref{fig:log-F densityplot}, log-$F(m,m)$ priors are
centered at zero and become increasingly concentrated as $m$ increases.
Consequently, larger values of $m$ lead to stronger shrinkage toward zero.
When $m=0$, the prior is flat and induces no shrinkage, which may be
appropriate for intercepts or covariates known \emph{a priori} to be strong
predictors. Because different covariates may warrant different degrees of
shrinkage, we allow a coefficient-specific value $m_k$ for each regression
coefficient $\beta_k$.

To calibrate $m_k$, we first choose a scientifically meaningful contrast
$x_k^u-x_k^l$ for covariate $x_k$. We then specify a plausible prior range
for the odds ratio associated with that contrast. Finally, $m_k$ is chosen
so that the log-$F$ prior assigns the desired prior probability
to that range. The log-odds ratio associated with the contrast is
\[
(x_k^u-x_k^l)\beta_k.
\]

For a binary covariate, the natural contrast is
$x_k^u-x_k^l=1$, giving the log-odds ratio $\beta_k$. For a continuous
covariate, the contrast should represent a meaningful change in exposure.
When a natural scientific contrast is unavailable, a scaling proposed by
\cite{Gelman2008} provides a convenient default. 
These authors proposed scaling continuous covariates by two standard deviations before assigning weakly informative priors to the resulting regression coefficients.
This is equivalent to using a contrast of approximately two standard deviations on the original
covariate scale, with
\[
x_k^u=\mu_k+\sigma_k,
\qquad
x_k^l=\mu_k-\sigma_k,
\]
so that the log-odds ratio is
\[
2\sigma_k\beta_k.
\]

Suppose that $(L,U)$ is regarded as a plausible central 95\% prior interval
for the log-odds ratio associated with the chosen contrast. We then select
$m_k$ so that
\[
P\!\left\{
L < (x_k^u-x_k^l)\beta_k < U
\right\}
=0.95.
\]
For a fixed plausible odds-ratio range, a larger covariate contrast implies
a narrower plausible range for $\beta_k$ itself and therefore requires a
more concentrated prior, corresponding to a larger value of $m_k$.

For example, with a binary covariate and contrast
$x_k^u-x_k^l=1$, $m=1$ gives a 95\% prior interval of approximately
$(-6.47,6.47)$ for the log-odds ratio $\beta_k$, giving odds ratios
between $1/648$ and $648$. This prior is essentially uninformative.
Increasing to $m=2$ narrows the 95\% prior interval to approximately
$(-3.66,3.66)$, giving odds ratios between $1/39$ and $39$.
This prior is weakly informative and produces stronger shrinkage toward zero.

Appendix~\ref{app:prior-calibration} describes the calibration procedure in
detail and provides R code for computing $m_k$ from a user-specified covariate
contrast and plausible odds-ratio range.

\subsection{Simulation Design}
\label{sec:simulations}

We conducted a simulation study to compare the frequentist operating characteristics of unpenalized conditional logistic regression, Firth's penalized method, and log-$F$--penalized conditional logistic regression. We evaluated the failure-to-converge rate of the iterative procedures, the bias and mean-squared error (MSE) of the exposure-effect estimator, the type-1 error rate and power of tests of the exposure effect, and the coverage of 95\% confidence intervals. The simulation settings were chosen to span a broad range of information available for estimating the exposure effect.

The simulation design varied the exposure type (binary or continuous), binary exposure prevalence ($1/20$, $1/10$, or $1/5$), exposure effect (0, 0.5, 1.0, or 1.5), case:control ratio (1:1 or 1:4), number of matched sets (10, 20, 30, 40, or 50), and number of nuisance covariates (0, 1, or 5). Varying the characteristics of the exposure, the matched-set design, and the complexity of the fitted regression model resulted in simulation settings ranging from relatively limited to relatively abundant information for estimating the exposure effect. These combinations yielded 120 configurations for continuous exposures and 360 configurations for binary exposures. Each configuration was evaluated using 10,000 independently generated datasets. All estimators were computed by iterative optimization with a maximum of 500 iterations.

For the log-$F(m,m)$ approach, the same log-$F(m,m)$ penalty was applied to the exposure and to all additional covariate effects.  
For binary exposures, we used $m=1$ and $m=2$ to specify uninformative and weakly informative priors, respectively. As discussed in Section~\ref{sec:choosem}, these correspond to 95\% prior ranges of $(-6.47,6.47)$ and $(-3.66,3.66)$.
For continuous exposures having an approximately standard normal distribution, we used $m=2.36$ and $m=5.62$ to specify uninformative and weakly informative priors. These values were obtained using the 
two-standard-deviation contrast and the calibration procedure described in Appendix~\ref{app:prior-calibration}, yielding
\[
P(-6.47<2\beta_k<6.47)=0.95
\]
for the uninformative prior and
\[
P(-3.66<2\beta_k<3.66)=0.95
\]
for the weakly informative prior. Throughout the simulation results, we use logFU and logFW to denote the uninformative and weakly informative log-$F(m,m)$ penalties, respectively.

\begin{figure}
    \centering
    \begin{tikzpicture}[node distance=2.5cm,
    every node/.style={
        circle,
        draw,
        minimum size=9mm
    }
]
    \node[] (H) {$H$};
    \node[right=of H] (E) {$E$};
    \node[right=of E] (D) {$D$};

    \draw[->, thick] (H) -- (E);
    \draw[->, thick] (E) -- (D);
    \draw[->, thick, bend left=35] (H) to (D);
\end{tikzpicture}
    \caption{Relationship between the hidden variable $H$ and the observed exposure $E$ and disease status $D$.}
    \label{fig:rships}
\end{figure}
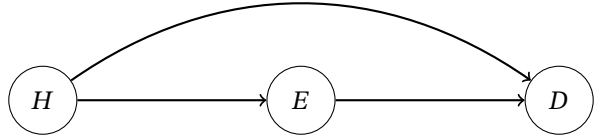

Parameter values were chosen to yield disease prevalences between 5\% and 10\%. This range represents an uncommon disease, as is typical in matched case--control studies, while also ensuring efficient simulation. The population variables were a latent confounder $H$, exposure $E$, nuisance covariates $Z$, and binary disease status $D$. We generated $H\sim N(0,1)$ and formed matched sets by conditioning on $H$. The covariates $Z$ were generated independently of $H$ and served only as nuisance variables in the fitted model. Figure~\ref{fig:rships} summarizes the assumed relationships among $H$, $E$, and $D$.

The conditional distribution of the exposure given the latent confounder was specified in one of two ways. For a binary exposure, we took
\[
(E\mid H=h)\sim\mathrm{Bernoulli}(p_h),
\qquad
\operatorname{logit}(p_h)=\alpha_0+h,
\]
where $\alpha_0$ was chosen to yield marginal exposure prevalences of $1/20$, $1/10$, and $1/5$. These prevalences represent increasingly sparse exposures, giving fewer exposed cases and therefore fewer informative matched comparisons. For a continuous exposure, we took
\[
(E\mid H=h)\sim N(h,1).
\]

Disease status was generated according to
\[
\operatorname{logit}\{P(D=1\mid E=e,H=h)\}
=
\beta_0+\beta_Ee+\beta_Hh,
\]
where $\beta_E\in\{0,0.5,1,1.5\}$ is the exposure effect, $\beta_H=2$ is the confounder effect, and $\beta_0=-5$ is the intercept. These values yielded empirical disease prevalences of approximately 5.4--8.4\% for continuous exposures and 3.4--5.8\% for binary exposures.

We generated 0, 1, or 5 nuisance covariates independently from a standard normal distribution. Increasing the number of nuisance covariates increased the number of regression coefficients that had to be estimated, thereby reducing the information available for estimation.

Because disease prevalence was low, cases were relatively rare. Therefore, for each sampled value $H=h$, we generated a temporary population of 10,000 individuals from the conditional distribution $(E,Z,D\mid H=h)$. One case and either $M=1$ or $M=4$ controls were then sampled to form a matched set. Repeating this procedure 10, 20, 30, 40, or 50 times yielded the desired number of matched sets. Increasing the number of matched sets or the number of controls per case increased the number of informative matched comparisons and therefore the information available for estimating the exposure effect.

For rare binary exposures, we occasionally obtained datasets with no exposed individuals. Such datasets provide no information about the exposure effect and would not be analyzed in practice. These datasets were discarded and regenerated to include at least one exposed individual.

\section{Data Application} \label{Data Application}

We illustrate the conditional logistic regression methods in two settings: a matched case-control study and a case-parent trio study. A case-parent trio can be represented as a matched case-control set by treating the affected child as the case and the three alternative offspring genotypes that could have arisen from the same parental alleles as matched pseudo-controls.

\subsection{Matched case-control study}

We first apply the conditional logistic-regression methods to the DES study of \cite{herbst1971adenocarcinoma}, which examined whether exposure to diethylstilbestrol (DES) during pregnancy was associated with subsequent development of vaginal cancer in daughters. The study included eight young women with vaginal cancer, each matched to four controls on birth date and hospital room type to reduce confounding by socioeconomic and hospital-related factors. The mothers of seven of the eight cases had received DES during pregnancy, whereas none of the mothers of the controls had received DES. DES exposure and maternal smoking status are summarized in Table~\ref{DES data}.

We use conditional logistic regression to estimate the effect of DES exposure while adjusting for maternal smoking. As noted earlier, DES exposure nearly separates the cases from the controls: seven of the eight cases were exposed, whereas none of the controls were exposed. When standard conditional logistic regression is fitted using the \texttt{clogit()} function in the \texttt{survival} package in \texttt{R}, a warning indicates that the DES coefficient has failed to converge. The corresponding conditional maximum likelihood estimate is therefore infinite.

\begin{table}[ht]
\centering
\caption{Data on cases and controls in the Herbst et al. study reconstructed from their Table 2.}
\normalsize
\begin{tabular}{cccc}
\hline
\textbf{DES} & \textbf{Maternal smoking} & \textbf{Cases} & \textbf{Controls}  \\ \hline
 yes & yes & 6 & 0  \\
 yes & no & 1 & 0  \\ 
 no & yes & 1 & 21 \\
 no & no & 0 & 11 \\ \hline
\end{tabular}%
\label{DES data}
\end{table}

\begin{figure}[ht]
\centering
\includegraphics[width=0.9\linewidth]{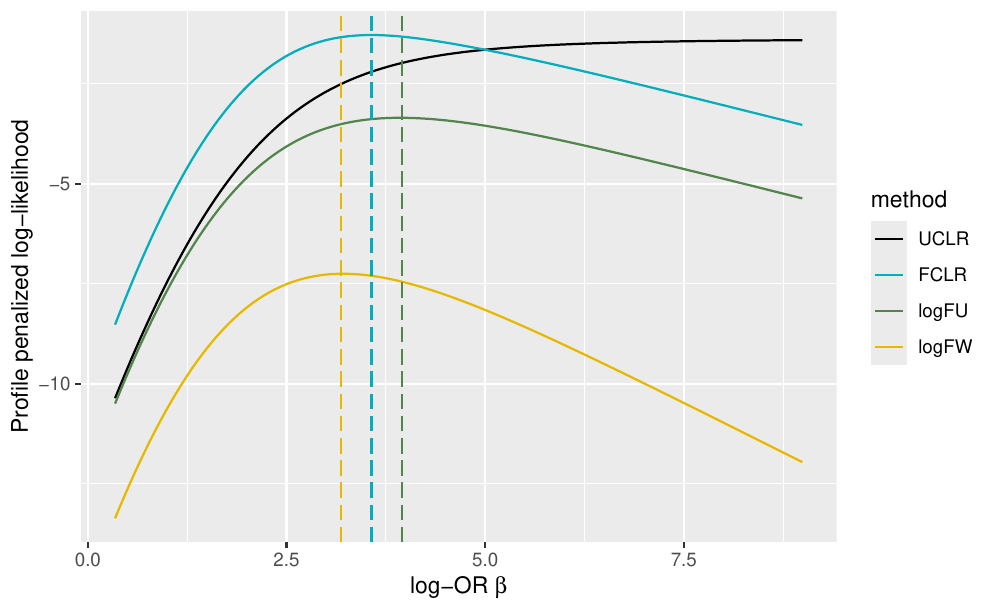}
\caption{Profile penalized log-likelihoods for the effect of exposure of DES, adjusting for the effect of maternal smoking.  Maximum profile penalized-likelihood estimators are indicated by vertical long-dashed lines.}
\label{PenLik_curve1}
\end{figure}

Estimated odds ratios for DES exposure adjusted for maternal smoking are shown in Table \ref{tab:des-results}. All methods yield estimated odds ratios and confidence intervals that strongly suggest an increased risk of vaginal cancer in DES-exposed daughters, with estimated odds ratios ranging from 24.8 to 50.9 depending on the amount of shrinkage. As expected, increasing $m$ shrinks the estimated odds ratio toward one and narrows the confidence intervals. To visualize the effect of penalization, Figure \ref{PenLik_curve1} displays the profile penalized log-likelihood, obtained by maximizing the penalized log-likelihood over the maternal smoking coefficient for each fixed value of the DES odds-ratio. The peak of each curve gives the corresponding penalized estimate, while the width of the curve reflects the uncertainty in that estimate.

\begin{table}
\centering
\caption{Penalized conditional logistic regression estimates for the effect of maternal DES exposure on their daughters' risk of vaginal cancer.}
\label{tab:des-results}
\normalsize
\begin{tabular}{lccc}
\hline
\textbf{Method} & \textbf{OR estimate} & \textbf{95\% CI} & \textbf{Std.\ err.} \\ \hline
Firth & 35.49 & (5.58, 4150.47) & 1.291 \\
log-$F(1,1)$ & 50.88 & (6.04, 6634.92) & 1.467 \\
log-$F(2,2)$ & 24.80 & (4.37, 465.99) & 1.074 \\ \hline
\end{tabular}
\end{table}

\subsection{Case-parent trio study}

We also applied the methods to data from 27 case-parent trios \citep{frayling1999parent}, from a study of the association between type 2 diabetes and the $Z{+}2$ allele at the \textit{GCK1} microsatellite locus (Table~\ref{case-parent trio data}). The data in the table are for trios with at least one heterozygous parent, with mating types denoted $x \times y$, where parental genotypes are coded by the number of $Z{+}2$ alleles carried (0, 1, or 2). Twenty-five of the trios are from the $0 \times 1$ mating type, two are from the $1 \times 1$ mating type, and none are from the $1 \times 2$ mating type. 
As no children in the study carry two copies of the $Z{+}2$ allele (genotype $g=2$), these data provide no information about the effect on disease risk of carrying two copies of the allele.
We therefore adopt a model with a single genotype relative risk (GRR), defined as the multiplicative change in disease risk for children carrying one copy of the $Z{+}2$ allele relative to children carrying no copies.

To estimate the GRR using conditional logistic regression, the data are reconstructed in the standard form for case-parent trio analyses (e.g. \citealt{shin2014data}). Each affected child is paired with matched pseudo-controls representing the alternative offspring genotypes consistent with the observed parental mating, and the resulting matched sets are stratified by parental mating type. The standard reconstruction is used except for heterozygous offspring from $1 \times 1$ matings, for which the transmitted parental alleles cannot be uniquely identified. The reconstruction of these ambiguous trios is described in the Supplementary Material. The conditional logistic regression model includes a single binary covariate indicating whether the offspring genotype contains a $Z{+}2$ allele.

Estimates and 95\% confidence intervals for the GRR under UCLR, FCLR, and the log-$F(1,1)$ and log-$F(2,2)$ penalties are summarized in Table~\ref{tab:trio-results}. As expected, the penalized methods lead to attenuated GRR estimates and narrower confidence intervals than UCLR. For all four methods, the estimated GRRs are close to one, and the corresponding confidence intervals include one, providing little evidence that the $Z{+}2$ allele at the \textit{GCK1} locus is associated with disease risk.


\begin{table}[ht]
\centering
\caption{Child's genotype in 27 case-parent trios}
\normalsize
\begin{tabular}{cccc}
\hline 
\multirow{2}{*}{\textbf{Parental mating type}} & \multicolumn{3}{c}{\textbf{Z+2 genotype ($g$)}} \\ 
 & \textbf{0} & \textbf{1} & \textbf{2} \\ \hline 
$0 \times 1$ & 10 & 15 & $-$ \\ 
$1 \times 1$ & 1 & 1 & 0 \\  \hline
\end{tabular}%
\label{case-parent trio data}
\end{table}

\begin{table}
\centering
\caption{Log-$F$--penalized estimates of the genotype relative risk (GRR) for one \textit{versus} zero copies of the Z+2 allele.}
\label{tab:trio-results}
\normalsize
\begin{tabular}{lccc}
\hline
Method & GRR estimate & 95\% CI & Std. err. \\
\hline
UCLR & 1.23 & (0.59,2.60) & 0.373 \\
FCLR & 1.22 & (0.60,2.55) & 0.373 \\
log-$F(1,1)$ & 1.22 & (0.60, 2.55) & 0.367 \\
log-$F(2,2)$ & 1.21 & (0.60, 2.50) & 0.361 \\
\hline
\end{tabular}
\end{table}


\section{Simulation Results}\label{sec:results}

Throughout, we use the shorthand logFU and logFW
for the log-$F(m,m)$ penalized approach with uninformative ($m=1$ for binary and 2.36 for continuous exposures) and weakly-informative ($m=2$ for binary and 5.62 for continuous exposures) priors, respectively. The shorthand UCLR and FCLR is used for unpenalized 
and Firth conditional logistic regression, respectively. Plots in the subsections below show estimates of operating characteristics and vertical error bars indicating their 95\% confidence intervals. In most cases the confidence intervals are completely obscured by the plotting symbols.

\subsection{Failure to converge}

For each simulation configuration, the non-convergence rate was estimated as the proportion of 10,000 simulated datasets for which the iterative fitting procedure failed to converge. Both log-$F$ methods converged for every simulated dataset with a binary exposure, and their non-convergence rates were below 0.1\% for every continuous-exposure configuration.

UCLR and FCLR both failed to converge in some simulation settings. The highest observed rate was 84.5\%, for UCLR under one-to-one matching with 10 matched sets, a continuous exposure, an exposure effect of 1.5, and five nuisance covariates. UCLR had a higher non-convergence rate than FCLR in 94.4\% of the 480 simulation configurations. In the remaining 5.6\%, however, FCLR had the higher rate.

\begin{figure}
\centering
\includegraphics[width=1.05\linewidth, height=0.8\linewidth]{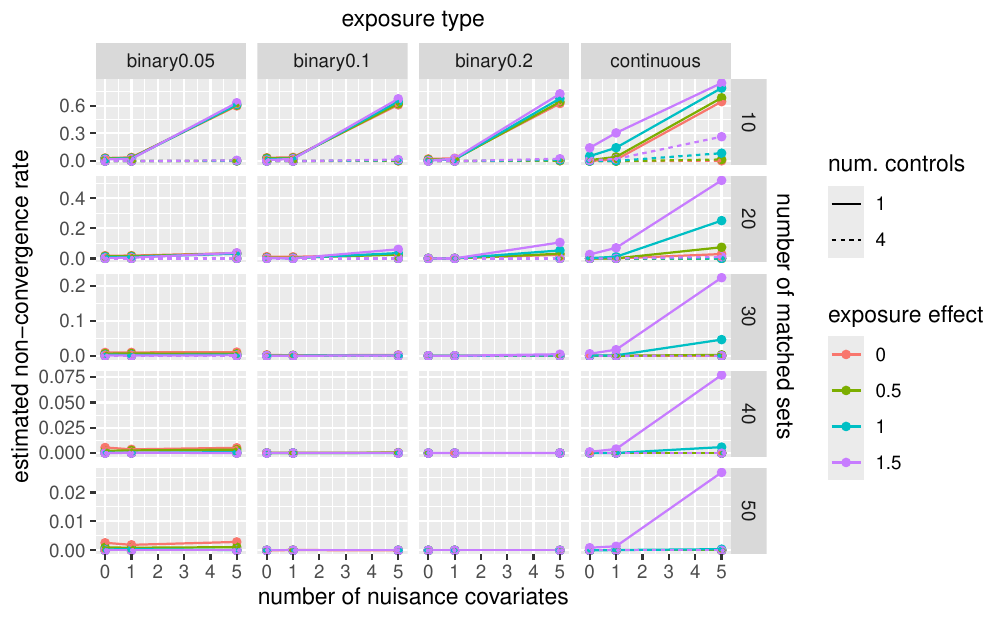}
\caption{Estimated non-convergence rates for UCLR based on 10,000 simulation replicates. Columns correspond to exposure type and prevalence, rows to the number of matched sets, and colours to the true exposure effect. Solid and dashed lines correspond to 1{:}1 and 1{:}4 matching, respectively. Note that the y-axis scales differ across rows of panels.}
\label{fig:nonconv}
\end{figure}

For both UCLR and FCLR, non-convergence tended to decrease as the number of matched sets increased and when four rather than one control was available per matched set. Non-convergence also tended to increase with the number of nuisance covariates and the magnitude of the exposure effect. The largest rates occurred in settings with the least information for estimating the exposure effect, particularly those with only 10 matched sets, one control per case, and five nuisance covariates. Non-convergence wa more pronounced for continuous than for binary exposures. Figure~\ref{fig:nonconv} shows the non-convergence rates for UCLR.

\subsection{Estimation of exposure effects}

We first evaluated the three methods for estimating the exposure effect. Bias and mean squared error (MSE) were used to assess point estimation, while coverage of nominal 95\% confidence intervals was used to assess interval estimation.
For clarity, the results
presented below are restricted to one-to-one matching, true exposure effects
of 0.5 and 1.5, and the two sample-size extremes of 10 and 50 matched sets.
Results for UCLR are omitted because their inclusion would have obscured the comparisons among the remaining methods.
Complete simulation results,
including those for one-to-four matching, intermediate exposure effects and
sample sizes, a binary exposure prevalence of 0.10, and a true exposure effect
of zero, are provided in the Supplementary Material.

Figure~\ref{fig:bias} shows the estimated bias of the exposure-effect estimator. 
Bias depended strongly on both the number of matched sets and the magnitude
of the exposure effect. With 50 matched sets, FCLR estimates were generally
close to unbiased and changed relatively little as the number of nuisance
covariates increased. In contrast, with only 10 matched sets and a true
exposure effect of 1.5, all three methods showed negative bias, although the
magnitude and pattern differed across methods. The negative bias was often
largest for logFW, while logFU tended to have less negative bias than FCLR
or logFW in these sparse settings.

For the log-$F$ methods, increasing the number of nuisance covariates tended
to move the bias in the positive direction. This pattern was most evident for
logFU with 50 matched sets and a true exposure effect of 1.5, for which the
bias became positive in several settings, particularly for a binary exposure
with prevalence 0.20 and for a continuous exposure. LogFW generally retained
more negative bias and therefore showed less of this positive shift. 
The tendency of logFU estimates to shift in the positive direction is consistent with 
its elevated type 1 
error rates discussed below.

\begin{figure}
\centering
\includegraphics[width=1.05\linewidth]{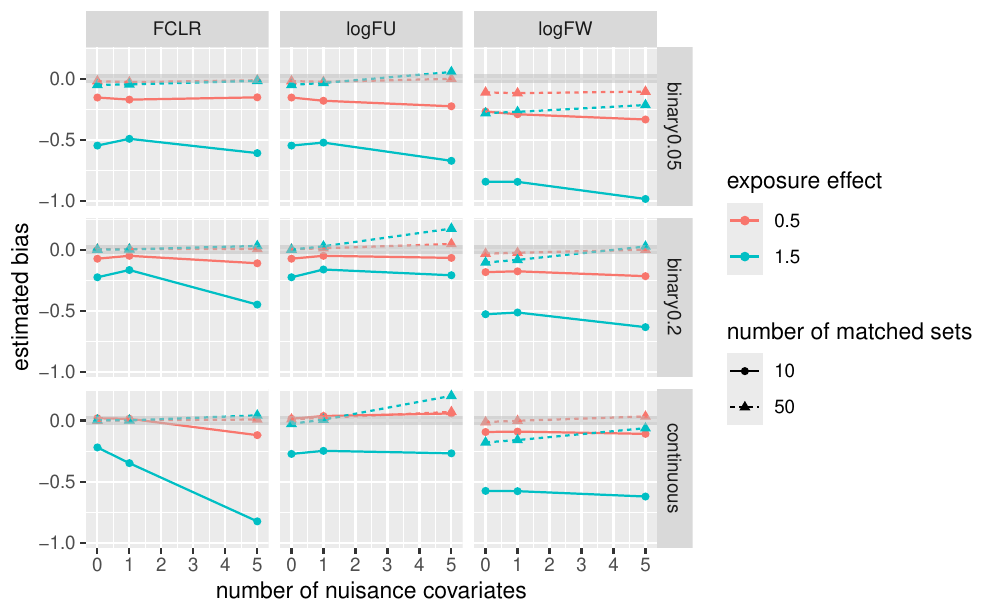}
\caption{Estimated bias for FCLR, logFU, and logFW under one-to-one
matching, based on 10,000 simulation replicates. Columns correspond to the
estimation method, and rows correspond to exposure type and, for binary
exposures, prevalence. The x-axis gives the number of nuisance covariates.
Colours indicate true exposure effects of 0.5 and 1.5, while plotting symbols
and line types indicate 10 and 50 matched sets.}
\label{fig:bias}
\end{figure}

\begin{figure}
\centering
\includegraphics[width=1.05\linewidth]{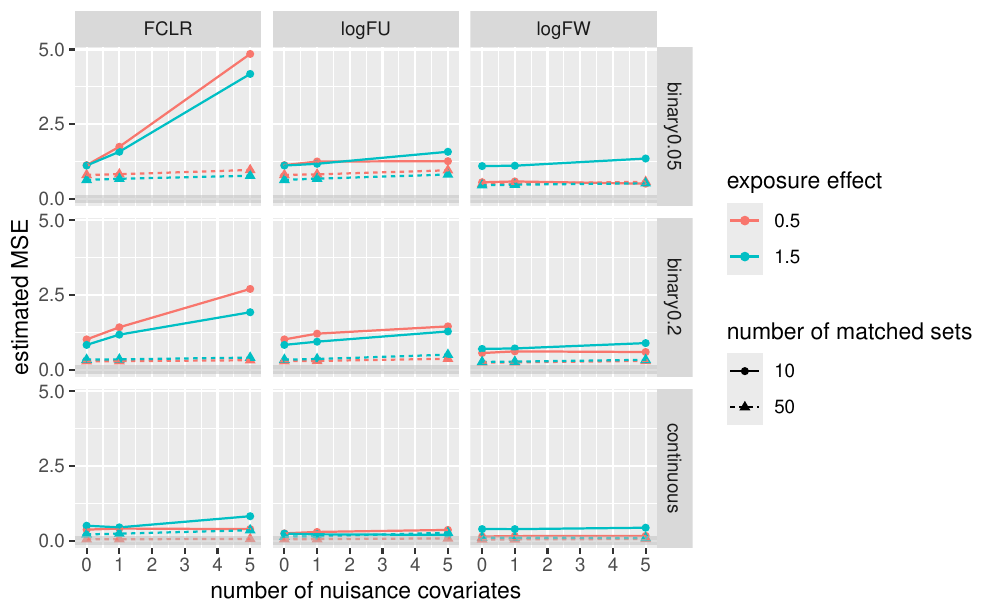}
\caption{Estimated MSE for FCLR, logFU, and logFW under one-to-one
matching, based on 10,000 simulation replicates. Columns correspond to the
estimation method, and rows correspond to exposure type and, for binary
exposures, prevalence. The x-axis gives the number of nuisance covariates.
Colours indicate true exposure effects of 0.5 and 1.5, while plotting symbols
and line types indicate 10 and 50 matched sets.}
\label{fig:MSE}
\end{figure}

Figure~\ref{fig:MSE} shows the estimated MSE of the exposure-effect
estimator. MSE depended strongly on the number of matched sets, decreasing
as the number of matched sets increased from 10 to 50. The
log-$F$ estimators generally had lower MSE than FCLR, with the largest 
differences occurring for 10 matched sets and
binary exposures. Among the log-$F$ estimators, logFW tended to have lower
MSE than logFU, although these differences were considerably smaller than
those between either log-$F$ estimator and FCLR.

The log-$F$ estimators also tended to have lower empirical variance than
FCLR, with the largest differences again occurring for 10 matched sets and
binary exposures (results not shown). The empirical variance results closely
paralleled the MSE results, indicating that the differences in MSE were
driven mainly by differences in sampling variability rather than squared
bias. Consistent with the MSE results, logFW tended to have lower empirical
variance than logFU.

\begin{figure}
\centering
\includegraphics[width=1.05\linewidth]{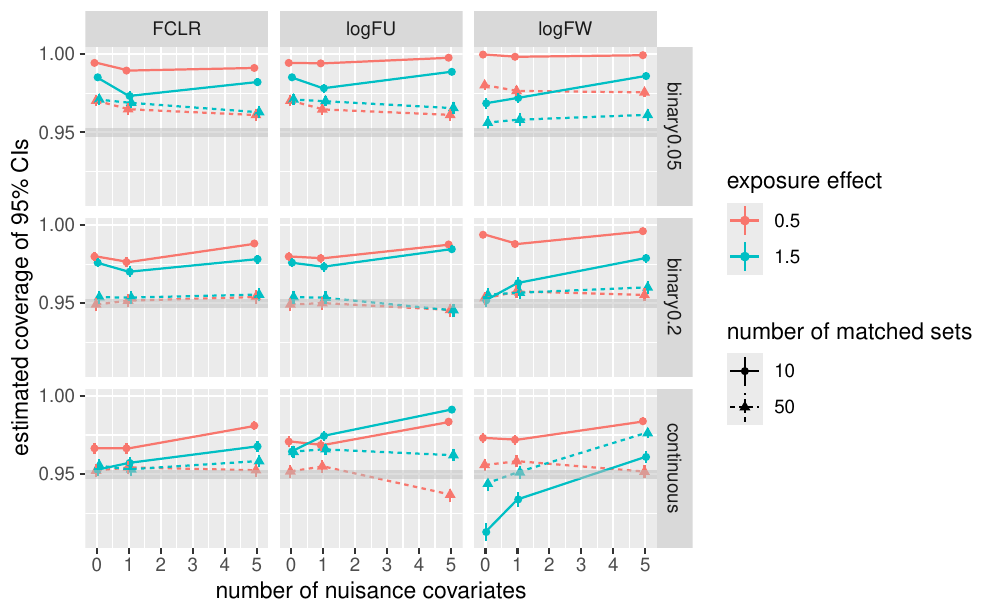}
\caption{Estimated coverage of 95\% confidence intervals for the exposure
effect for FCLR, logFU, and logFW under one-to-one matching, based on
10,000 simulation replicates. Columns correspond to the estimation method,
and rows correspond to exposure type and, for binary exposures, prevalence.
The x-axis gives the number of nuisance covariates. Colours indicate true
exposure effects of 0.5 and 1.5, while plotting symbols and line types
indicate 10 and 50 matched sets. 95\% confidence intervals for the estimates are 
indicated by vertical lines that are often
obscured by the plotting symbols.}
\label{fig:cover}
\end{figure}

Figure~\ref{fig:cover} shows the estimated coverage of the 95\% confidence
intervals for the exposure effect. Coverage was for the most part close to or above
the nominal 95\% level for all three methods. The most notable departures
from nominal coverage occurred for logFU with a continuous exposure, a true
exposure effect of 0.5, 50 matched sets, and five nuisance covariates, and
for logFW with a continuous exposure, a true exposure effect of 1.5, 10
matched sets, and zero or one nuisance covariate. These configurations also
corresponded to positive bias for logFU and negative bias for
logFW, respectively, suggesting that bias contributed to the observed
undercoverage. Bias alone, however, did not determine coverage. For example,
FCLR and logFW both had substantial negative bias for a continuous exposure
with a true exposure effect of 1.5, 10 matched sets, and five nuisance
covariates, yet their confidence intervals retained at least nominal
coverage.

\subsection{Testing of exposure effects}

We next evaluated the three methods for testing the null hypothesis of no exposure effect. 
Type 1 error was assessed using the empirical type 1 error rate under the null hypothesis, 
while power was assessed using the empirical power under nonzero exposure effects. 
For clarity, the type 1 error results presented
below are restricted to one-to-one matching and a true exposure effect of
zero, while the power results are restricted to one-to-one matching and true
exposure effects of 0.5 and 1.5. In both cases, we show only the two
sample-size extremes of 10 and 50 matched sets. Results for UCLR are omitted
because their inclusion would have obscured the comparisons among the
remaining methods. Complete simulation results, including those for
one-to-four matching, intermediate exposure effects and sample sizes, and a
binary exposure prevalence of 0.10, are provided in the Supplementary
Material.

Figure~\ref{fig:t1e} shows the empirical type 1 error rates for tests of the
exposure effect. FCLR and logFW largely maintained type 1 error rates close
to the nominal 5\% level across the simulation settings shown, whereas logFU
showed mild inflation in several settings. The inflation was most apparent
for continuous exposures and became more pronounced as the number of nuisance
covariates increased. Increasing the number of matched sets brought
the empirical type 1 error rate closer to the nominal level, although logFU remained
mildly anticonservative with five nuisance covariates even with 50 matched
sets. This pattern is consistent with the tendency of logFU estimates to
shift in the positive direction observed in Figure~\ref{fig:bias}.

\begin{figure}
\centering
\includegraphics[width=1.05\linewidth]{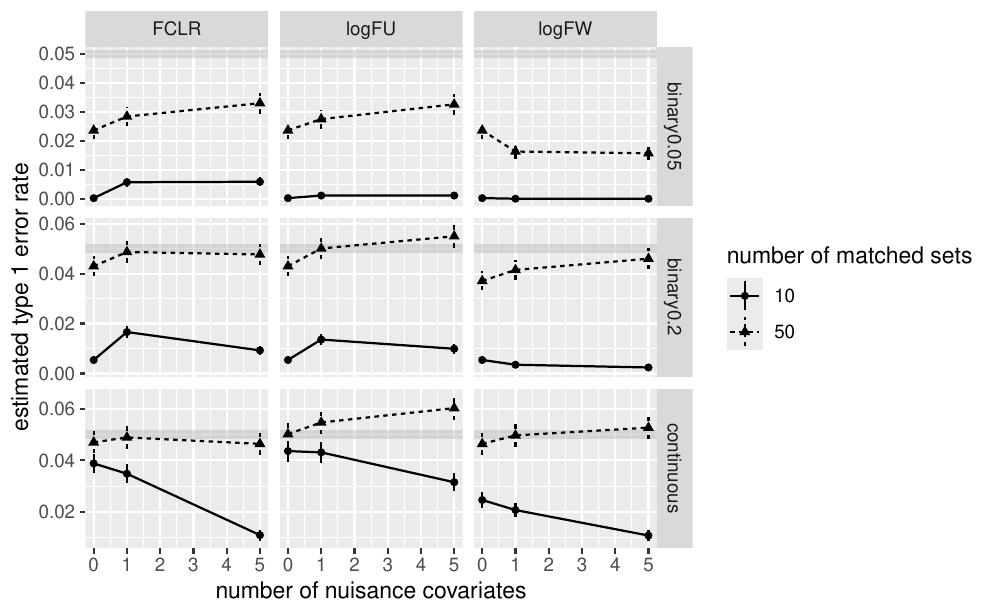}
\caption{Estimated type 1 error rates for FCLR, logFU, and logFW tests of the
exposure effect under one-to-one matching, based on 10,000 simulation
replicates. Columns correspond to the estimation method, and rows correspond
to exposure type and, for binary exposures, prevalence. The x-axis gives the
number of nuisance covariates. Plotting symbols and line types indicate 10
and 50 matched sets. 95\% confidence intervals for the estimates are 
indicated by vertical lines that are occasionally
obscured by the plotting symbols. The y-axis limits differ across rows.}
\label{fig:t1e}
\end{figure}

\begin{figure}
\centering
\includegraphics[width=1.0\linewidth]{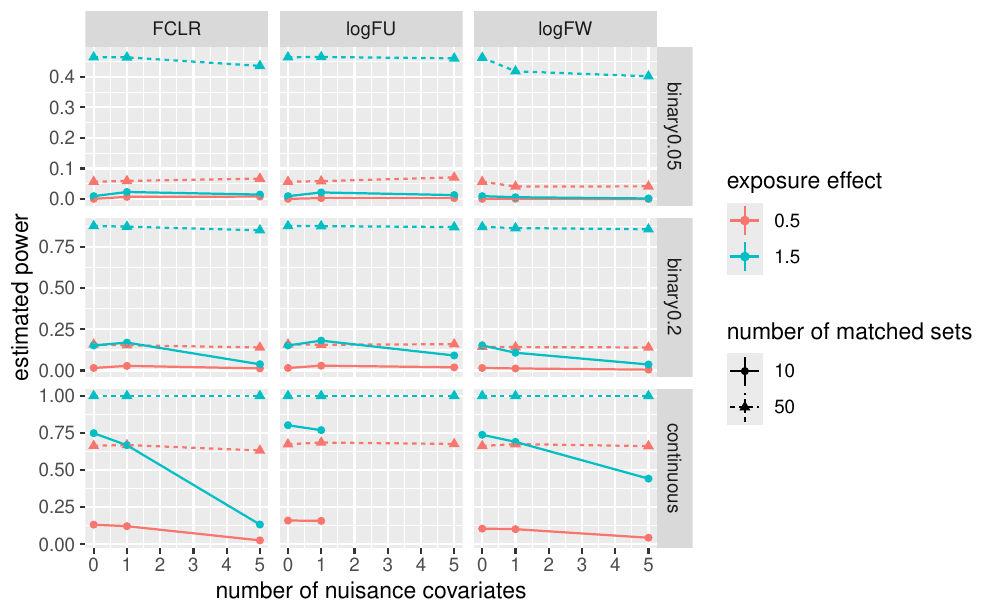}
\caption{Estimated power for FCLR, logFU, and logFW tests of the exposure
effect under one-to-one matching, based on 10,000 simulation replicates.
Columns correspond to the estimation method, and rows correspond to exposure
type and, for binary exposures, prevalence. The x-axis gives the number of
nuisance covariates. Colours indicate true exposure effects of 0.5 and 1.5,
while plotting symbols and line types indicate 10 and 50 matched sets. The
y-axis limits differ across rows. Power results for logFU with 10 matched
sets, a continuous exposure, and five nuisance covariates are not shown
because the corresponding test did not control the type 1 error rate. 
The y-axis limits differ across rows.}
\label{fig:power}
\end{figure}

Figure~\ref{fig:power} shows the empirical power to detect the exposure
effect. Power depended most strongly on the magnitude of the exposure effect
and the number of matched sets, and tended to be higher for continuous
exposures and for binary exposures with prevalence 0.20 than for binary
exposures with prevalence 0.05. Increasing the number of nuisance covariates
tended to reduce power, particularly with only 10 matched sets. In comparison with 
these effects of the simulation design, differences in power among FCLR, logFU, and logFW were relatively modest.
LogFU sometimes had slightly higher
empirical power, but it also had inflated type 1 error rates in some settings. Among
the methods that adequately controlled type 1 error, FCLR and logFW had
broadly similar power overall, with neither method consistently more powerful
across the simulation settings shown.

\section{Discussion}
\label{Discussion}

We have proposed a log-$F(m,m)$ penalized conditional logistic regression
estimator for sparse matched and stratified data. The method extends
log-$F$ penalization from ordinary logistic regression to conditional
likelihoods and can be implemented using simple data augmentation together
with standard conditional logistic regression software. For matched pairs
with a binary exposure, the proposed estimator reduces to the classical
Haldane and Laplace corrections as special cases, providing a
natural generalization of these longstanding corrections to more general
matched and stratified designs.

Our simulations confirmed the well-known difficulties of unpenalized
conditional logistic regression in sparse-data settings. When the number
of matched sets was small or the number of nuisance covariates was large,
the conditional maximum likelihood estimator frequently provided unstable
estimates or failed to converge. Both Firth-penalized conditional logistic
regression and log-$F$ penalization largely eliminated these problems,
providing finite estimates even in settings where the conditional maximum
likelihood estimator failed to exist or was numerically unstable.

Our evaluation focused on frequentist operating characteristics,
including bias, mean-squared error, confidence-interval coverage,
type~1 error, and power. These criteria facilitate direct comparison of the log-$F$ penalized approach
with existing penalized likelihood methods such as Firth's approach.
Although the log-$F$ penalty may be viewed as arising from a prior
distribution, our objective was not to apply a fully Bayesian
procedure, but rather to use an interpretable penalty that yields
good frequentist operating characteristics in sparse matched and stratified
studies.

A practical challenge in applying any shrinkage method is determining the
appropriate amount of shrinkage. A useful feature of the log-$F$
approach is that this choice is controlled by a single interpretable
parameter, $m$. Rather than selecting an abstract tuning parameter,
investigators instead specify a plausible range of odds ratios for a
scientifically meaningful covariate contrast. The corresponding value of
$m$ is then obtained automatically from this specification. Building on
the recommendations of \cite{greenland2015penalization}, this calibration
provides a transparent connection between prior scientific knowledge and
the amount of shrinkage applied during estimation. For binary exposures,
for example, a log-$F(1,1)$ prior places approximately 95\% of its mass
on odds ratios between $1/648$ and $648$, a range that would be
regarded as essentially uninformative. Increasing $m$ corresponds to
specifying progressively narrower plausible odds-ratio ranges and hence
stronger shrinkage toward the null.

For continuous exposures, investigators may calibrate the prior using
scientifically meaningful exposure contrasts. In the absence of more
specific prior information, we 
adapted the two-standard-deviation scaling rule of \cite{Gelman2008}.
This rule yields log-$F(2.36,2.36)$ and
log-$F(5.62,5.62)$ priors as analogues of the uninformative log-$F(1,1)$
and weakly informative log-$F(2,2)$ priors, respectively, for binary
exposures. Reassuringly, simulation results for these calibrated priors
were consistent with those obtained for binary exposures, supporting the
two-standard-deviation rule as a practical default when 
no more scientifically meaningful exposure contrast is available.

Perhaps the most surprising finding of our simulations was that, once the penalties could be distinguished, the preferred level of shrinkage within the log-$F$ family depended on whether the information available for estimating the exposure effect was limited by the number of matched sets, the exposure prevalence, or the number of nuisance covariates. The preferred level of shrinkage was relatively insensitive to the number of matched sets and the exposure prevalence, but increasing the number of nuisance covariates shifted the preference toward the weakly informative log-$F$ penalty.
With more matched sets and only a few nuisance covariates, the weaker, uninformative log-$F$ prior maintained satisfactory type~1 error control and confidence-interval coverage and had slightly greater power than the weakly informative log-$F$ prior and Firth penalization. When several nuisance covariates were included, however, the weaker penalty no longer adequately controlled the type~1 error rate and, in some settings, confidence-interval coverage. In contrast, the weakly informative log-$F$ prior maintained type~1 error rates close to the nominal level with power and confidence-interval coverage comparable to Firth penalization while achieving lower mean-squared error. This lower mean-squared error is consistent with the motivation of \cite{greenland2015penalization} for log-$F$ penalization. Because the strength of the log-$F$ penalty can be calibrated, it can achieve a more favourable bias-variance tradeoff than the fixed amount of shrinkage induced by Firth's penalty.  Taken together, these results favour the weakly informative calibration for point estimation and for inference in settings with several nuisance covariates, whereas either log-$F$ calibration appears reasonable for inference when only a few nuisance covariates are included.


The choice between the log-$F$ approach and Firth-penalized conditional
logistic regression is not based solely on their operating
characteristics. Firth's method is based on the Jeffreys prior and
therefore depends on the observed information matrix, whereas the
log-$F$ approach specifies the prior independently of the observed
data. Consequently, the amount of shrinkage is determined entirely by
scientifically plausible effect sizes rather than features of the
observed dataset. This may be advantageous when prior knowledge
regarding plausible odds ratios is available.

Several limitations should be noted. Our simulations considered only
$1\!\!:\!\!M$ matched designs with a single exposure of interest and
independent nuisance covariates. 
Additional work is needed to investigate the frequentist
operating characteristics of log-$F$ penalization when multiple exposures are of
simultaneous scientific interest, when covariates are highly
correlated or include interaction effects, and in higher-dimensional regression
models. Finally, our observation that the number of nuisance covariates
strongly influences the appropriate degree of shrinkage was unexpected
and deserves further theoretical investigation.

In summary, log-$F$ penalized conditional logistic regression provides a
simple and practical approach for sparse matched and stratified studies.
The method yields finite estimates under separation, can be implemented
through straightforward data augmentation using standard software, and
performs competitively with Firth-penalized conditional logistic
regression. Perhaps its greatest practical advantage is that the degree
of shrinkage can be specified in terms of scientifically meaningful
odds-ratio ranges rather than an abstract tuning parameter. We therefore
recommend log-$F$ penalization as a useful addition to the toolbox for
sparse-data analysis in matched and stratified studies.

\section*{Acknowledgements}

The authors would like to thank Georg Heinze for guidance on using the \texttt{coxphf()} function to implement Firth-penalized conditional logistic regression.

\section*{Software Availability}
The R code used to analyze the data and conduct the simulation study are available on GitHub at \url{https://github.com/SFUStatgen/logF_CLR}. The specific version of the code used to prepare this manuscript is in release 1.0.0, available at \url{https://github.com/SFUStatgen/logF_CLR/releases/tag/1.0.0}. Simulation results are archived on Zenodo at \url{https://zenodo.org/records/21708371}, reference number 10.5281/zenodo.21708371.

\appendix

\section{Haldane and Laplace methods}
\label{HLlogF}

Here we show that the Haldane and Laplace corrections for matched-pairs data
with a single binary exposure are special cases of log-$F$-penalized
conditional logistic regression.

Matched-pairs data with a single binary exposure may be summarized by the
$2\times2$ table of case and matched-control exposure status shown in
Table~\ref{tab:matchedpairs}. The conditional log-likelihood from
equation~(\ref{cond_log_lkhd}) is
\begin{equation}
l(\beta_1)
=
b\{\beta_1-\log(1+\exp(\beta_1))\}
-
c\log(1+\exp(\beta_1)),
\label{eqn:lkhdpairs}
\end{equation}
where only the discordant pairs contribute to the likelihood.
Differentiating (\ref{eqn:lkhdpairs}) and solving the score equation gives the
maximum likelihood estimator
\[
\hat{\beta}_1=\log(b/c)
\]
\citep{BreslowDay80}. Consequently, the estimator is infinite whenever
$b=0$ or $c=0$.

\begin{table}
\centering
\caption{Summary table of exposure status for matched pairs.}
\normalsize
\begin{tabular}{cccc}
        & & \multicolumn{2}{c}{Control} \\
        & & exposed & unexposed \\
Case & exposed & $a$ & $b$ \\
     & unexposed & $c$ & $d$
\end{tabular}
\label{tab:matchedpairs}
\end{table}

The Haldane and Laplace corrections avoid infinite estimates by adding
$m/2$ to each discordant cell, with $m=1$ and $m=2$, respectively.
Equivalently, they may be viewed as the maximum likelihood estimator obtained
from an augmented dataset. The corresponding log-likelihood is
\begin{align}
l^a(\beta_1)
&=
(b+m/2)\{\beta_1-\log(1+\exp(\beta_1))\}
-
(c+m/2)\log(1+\exp(\beta_1))
\nonumber\\
&=
l(\beta_1)
+
\frac{m}{2}
\left\{
\beta_1
-
2\log(1+\exp(\beta_1))
\right\}.
\label{eqn:penlkhdpairs}
\end{align}

The second term in (\ref{eqn:penlkhdpairs}) arises from the data
augmentation and, up to an additive constant, is the logarithm of a
log-$F(m,m)$ density for $\beta_1$ (see equation \ref{eqn:penaltycontribk}). Thus, augmenting the discordant cells by
$m/2$ is equivalent to fitting a log-$F(m,m)$-penalized conditional logistic
regression model. Haldane's and Laplace's corrections therefore correspond to
log-$F(1,1)$ and log-$F(2,2)$ penalization, respectively.

\section{Calibration of $m$}
\label{app:prior-calibration}

Section~\ref{sec:choosem} proposes calibrating $m$ by first choosing a scientifically meaningful covariate contrast 
$x_u-x_l$ and then specifying a plausible prior range for the odds ratio. This appendix describes the 
resulting calculation and provides R code for its implementation.

For a binary covariate, the natural contrast is $x_u-x_l=1$.
For a continuous covariate, the contrast should represent a meaningful change in exposure.
When no natural scientific contrast is available, a convenient default is the scaling proposed by \cite{Gelman2008}, 
which is equivalent to calibrating effects over a contrast of approximately two standard deviations on the original covariate scale, i.e.,
$x_u-x_l = 2\sigma_x$. More generally, the user may specify any contrast $x_u-x_l$ directly. 
For any chosen contrast, the log-odds ratio is
\[
(x_u-x_l)\beta.
\]

One can specify a plausible prior range for the odds ratio associated with
the contrast $x_u-x_l$ by choosing an upper limit
$\mathrm{OR}_{\max}$ and a prior probability $1-\alpha$ that the odds ratio
lies in the interval
\[
\left(1/\mathrm{OR}_{\max},\mathrm{OR}_{\max}\right).
\]
The upper limit corresponds to the maximum plausible log-odds ratio,
\[
(x_u-x_l)\beta_{\max}
=
\log(\mathrm{OR}_{\max}),
\]
or, equivalently,
\[
\beta_{\max}
=
\frac{\log(\mathrm{OR}_{\max})}
     {x_u-x_l}.
\]
Thus, for a fixed plausible odds-ratio range, larger covariate contrasts imply smaller values of
$\beta_{\max}$, requiring a more concentrated prior and hence a larger value of $m$.
The required value of $m$ satisfies
\[
P(-\beta_{\max}<B<\beta_{\max})=1-\alpha,
\qquad
B\sim \log\text{-}F(m,m).
\]
Equivalently, because $Y=\exp(B)\sim F(m,m)$,
\[
Q_{F(m,m)}(\alpha/2)
=
\exp(-\beta_{\max}),
\]
which can be solved numerically for $m$.

The following R code implements the calibration procedure. The function
\texttt{choosem()} computes the calibrated value of $m$ from a
user-specified upper limit $\mathrm{OR}_{\max}$, covariate contrast
(default 1), and prior probability that the odds ratio lies in
$(1/\mathrm{OR}_{\max},\mathrm{OR}_{\max})$ (default 0.95).

\bigskip
\small

\begin{verbatim}
## Choose m by specifying
##   ORmax     = upper limit of the plausible odds-ratio range
##   contrast  = covariate contrast x_u - x_l
##   level     = prior probability assigned to that range
##
## The function returns m satisfying
##
## P(-beta_max < B < beta_max) = level,
##     B ~ log-F(m,m),
##
## where beta_max = log(ORmax)/contrast.

choosem <- function(ORmax, contrast = 1, level = 0.95) {
  beta.max <- log(ORmax) / contrast
  solvem(exp(beta.max), level)
}

solvem <- function(ORmax, level = 0.95) {
  ORmin <- 1 / ORmax
  f <- function(m)
    ORmin - qf((1 - level)/2, m, m)
  uniroot(f, interval = c(0.01, 100))$root
}

# Example calibrations:
choosem(648, contrast = 1)   # returns m=1
choosem(648, contrast = 2)   # returns m=2.36
choosem(39,  contrast = 1)   # returns m=2
choosem(39,  contrast = 2)   # returns m=5.62
\end{verbatim}

\bigskip
\normalsize

The calibrated value of $m$ need not be an integer. If the data-augmentation implementation requires an even integer value, one may round $m$ to a nearby even integer. Rounding upward produces slightly greater shrinkage of the estimate of the conditional logistic regression coefficient toward zero, whereas rounding downward produces slightly less shrinkage.

\bibliographystyle{plainnat}
\bibliography{sn-bibliography}

\end{document}